\documentclass{PoS}
\usepackage{amsmath,amssymb,epsfig,fleqn,url,psboxit,color,here,graphics,graphicx,rotating,multirow}
\usepackage{wrapfig}

\title{Transverse spin asymmetries at COMPASS: \\beyond Collins and Sivers effects}

\ShortTitle{Transverse spin asymmetries at COMPASS}

\author{\speaker{Bakur Parsamyan}\\
        Dipartimento di Fisica Generale, Universit\`a di Torino, Torino, Italy\\
        INFN, Sezione di Torino, Via P. Giuria 1, I-10125 Torino, Italy\\
        E-mail: \email{bakur.parsamyan@cern.ch}}

\abstract{One of the important objectives of the COMPASS experiment (SPS, CERN) \cite{Abbon:2007pq} is the exploration of the transverse spin structure of the nucleon via spin dependent azimuthal asymmetries in single-hadron production in deep inelastic scattering of polarized leptons off transversely polarized targets. For this purpose a series of measurements were made in COMPASS, using 160 GeV/c longitudinally polarized muon beam and transversely polarized $^6LiD$ (in 2002, 2003 and 2004) and $NH_3$ (in 2007 and 2010) targets.	
	In the past few years considerable theoretical interest and experimental efforts were focused on the study of Collins and Sivers transverse spin asymmetries. The experimental results obtained so far play an important role in the general understanding of the three-dimensional nature of the nucleon in terms of transverse momentum dependent parton distribution functions.	
	In addition to these two measured leading-twist effects, the SIDIS cross-section includes six more target transverse spin dependent azimuthal asymmetries, which have their own well defined leading or higher-twist interpretation in terms of QCD parton model.

COMPASS preliminary results for these six "beyond Collins and Sivers" asymmetries, obtained from transversely polarized deuteron and proton data have been presented at the previous conferences \cite{Parsamyan:2007ju} - \cite{Parsamyan:2013ug}. In this review we focus on the results obtained with the last "proton-2010" data sample.
}

\FullConference{XXI International Workshop on Deep-Inelastic Scattering and Related Subject -DIS2013,\\
		22-26 April 2013\\
		Marseilles,France}

\begin{document}
\section{\label{sec:intro}Introduction}

\begin{wrapfigure}{r}{0.42\textwidth}
  \begin{center}
    \includegraphics[width=0.42\textwidth]{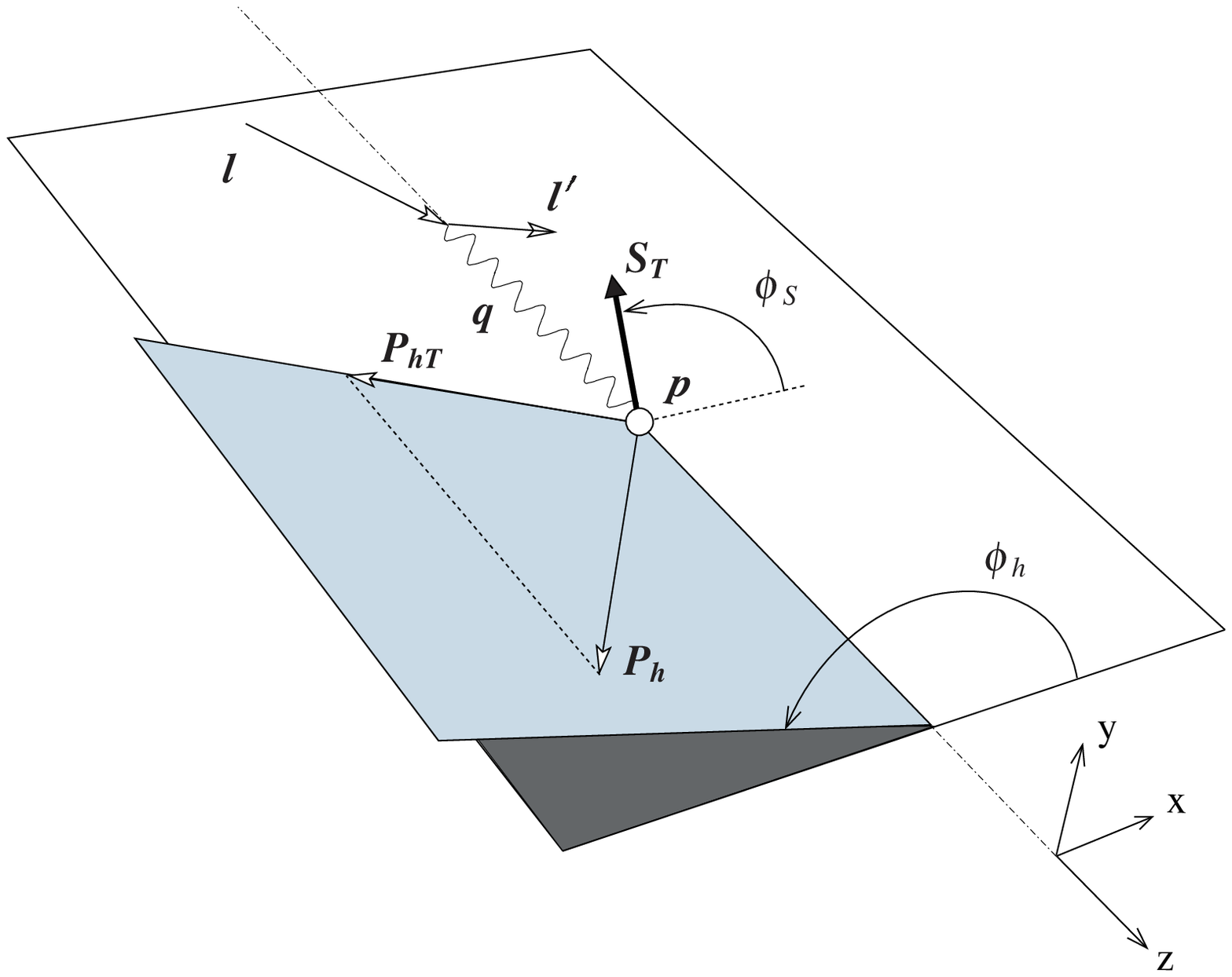}
    \includegraphics[width=0.25\textwidth]{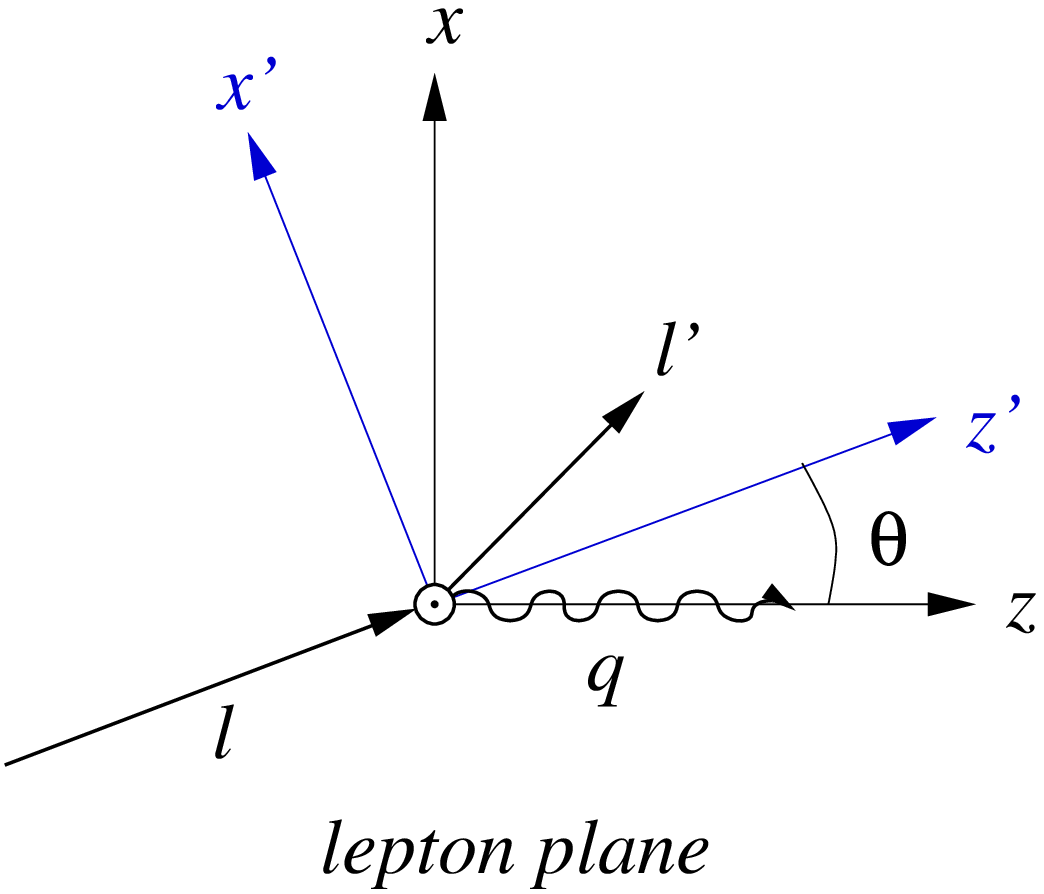}
  \end{center}
  \caption{SIDIS framework (top) and scheme of $lp\rightarrow\gamma^*p$ transition (bottom)}
  \label{fig:SIDIS}
\end{wrapfigure}

Measurement of the azimuthal asymmetries in SIDIS of leptons off polarized proton and deuteron targets is a powerful tool to study the spin structure of the nucleon. Schematic view of the SIDIS framework and some notations and definitions adopted in this letter \footnote{The notations applied in this contribution are equivalent to those adopted in \cite{Kotzinian:1994dv},\cite{Bacchetta:2006tn}, \cite{Parsamyan:2007ju}-\cite{Parsamyan:2013ug} and $\theta$ is the angle between $\gamma^*$-direction and initial lepton momenta (Fig.\ref{fig:SIDIS})} such as system of axes, azimuthal angles, target polarization, etc. are presented in Fig.~\ref{fig:SIDIS}.
Here the target transverse polarization ($S_T$) is defined relative to the virtual photon momentum direction (which is a most natural basis from the theory point of view) while, in experiment transverse polarization ($P_T$) is defined relative to the beam (incoming lepton) direction. As it was demonstrated in \cite{Kotzinian:1994dv}-\cite{Diehl:2005pc} this difference, in particular, influences azimuthal distributions in the final state. After the appropriate conversions, the model-independent expression for the SIDIS cross-section for transversely (w.r.t. lepton beam) polarized target can be re-written in the following way \cite{Kotzinian:1994dv}-\cite{Diehl:2005pc}:
{\footnotesize
\begin{eqnarray}
\label{eq:x_sec_mod}
  && \hspace*{-0.0cm}\frac{{d\sigma }}{{dxdydzdP_{hT}^2d{\varphi _h}d{\varphi _S}}} = \\
 &&\hspace*{0.1cm}  \left[ {\frac{{\cos \theta }}{{1 - {{\sin }^2}\theta {{\sin }^2}{\varphi _S}}}} \right]  \left[ {\frac{\alpha }{{xy{Q^2}}}\frac{{{y^2}}}{{2\left( {1 - \varepsilon } \right)}}\left( {1 + \frac{{{\gamma ^2}}}{{2x}}} \right)} \right]  \left( {{F_{UU,T}} + \varepsilon {F_{UU,L}}} \right) \times \nonumber
\end{eqnarray}
\[\begin{gathered}[h!]\hspace*{-0.0cm} \left( \begin{gathered}
  1 + \cos {\varphi _h}  \sqrt {2\varepsilon \left( {1 + \varepsilon } \right)} A_{UU}^{\cos {\varphi _h}} + \cos {2{\varphi _h}}  \varepsilon  A_{UU}^{\cos {2{\varphi _h}}} +   \lambda \sin {\varphi _h}  \sqrt {2\varepsilon \left( {1 - \varepsilon } \right)} A_{LU}^{\sin {\varphi _h}} +
  \hfill\\
  \frac{{\text{P}}_{\text{T}}}{{{\sqrt {1 - {{\sin }^2}\theta {{\sin }^2}{\varphi _S}} }}} \left[ \begin{gathered}
  \sin {\varphi _S}  \left( \cos \theta {\sqrt {2\varepsilon \left( {1 + \varepsilon } \right)}  A_{UT}^{\sin {\varphi _S}}} \right) +  \hfill \\
  \sin \left( {{\varphi _h} - {\varphi _S}} \right)  \left( {\cos \theta A_{UT}^{\sin \left( {{\varphi _h} - {\varphi _S}} \right)} + \frac{1}{2}\sin \theta\sqrt {2\varepsilon \left( {1 + \varepsilon } \right)} A_{UL}^{\sin {\varphi _h}}} \right) +  \hfill \\
  \sin \left( {{\varphi _h} + {\varphi _S}} \right)  \left(\cos \theta  {\varepsilon A_{UT}^{\sin \left( {{\varphi _h} + {\varphi _S}} \right)} + \frac{1}{2}\sin \theta \sqrt {2\varepsilon \left( {1 + \varepsilon } \right)} A_{UL}^{\sin {\varphi _h}}} \right) +  \hfill \\
  \sin \left( {2{\varphi _h} - {\varphi _S}} \right)  \left( \cos \theta{\sqrt {2\varepsilon \left( {1 + \varepsilon } \right)}  A_{UT}^{\sin \left( {2{\varphi _h} - {\varphi _S}} \right)} + \frac{1}{2}\sin \theta \varepsilon A_{UL}^{\sin 2{\varphi _h}}} \right) +  \hfill \\
  \sin \left( {3{\varphi _h} - {\varphi _S}} \right)  \left(\cos \theta {\varepsilon  A_{UT}^{\sin \left( {3{\varphi _h} - {\varphi _S}} \right)}} \right) +
    \sin \left( {2{\varphi _h} + {\varphi _S}} \right)  \left( {\frac{1}{2}\sin \theta \varepsilon A_{UL}^{\sin 2{\varphi _h}}} \right) \hfill \\
\end{gathered}  \right] +  \hfill \\
  \frac{{{\text{P}}_{\text{T}}}{\lambda}}{{{\sqrt {1 - {{\sin }^2}\theta {{\sin }^2}{\varphi _S}} }}} \left[ \begin{gathered}
  \cos {\varphi _S}  \left( \cos \theta {\sqrt {2\varepsilon \left( {1 - \varepsilon } \right)} A_{LT}^{\cos {\varphi _S}} +  \sin \theta \sqrt {\left( {1 - {\varepsilon ^2}} \right)} {A_{LL}}} \right) +  \hfill \\
  \cos \left( {{\varphi _h} - {\varphi _S}} \right)  \left( \cos \theta {\sqrt {\left( {1 - {\varepsilon ^2}} \right)} A_{LT}^{\cos \left( {{\varphi _h} - {\varphi _S}} \right)} + \frac{1}{2} \sin \theta \sqrt {2\varepsilon \left( {1 - \varepsilon } \right)} A_{LL}^{\cos {\varphi _h}}} \right) +  \hfill \\
    \cos \left( {2{\varphi _h} - {\varphi _S}} \right)  \left( \cos \theta {\sqrt {2\varepsilon \left( {1 - \varepsilon } \right)} A_{LT}^{\cos \left( {2{\varphi _h} - {\varphi _S}} \right)}} \right) + \hfill \\
  \cos \left( {{\varphi _h} + {\varphi _S}} \right)  \left( {\frac{1}{2}\sin \theta \sqrt {2\varepsilon \left( {1 - \varepsilon } \right)} A_{LL}^{\cos {\varphi _h}}} \right) \hfill
\end{gathered}  \right] \hfill
\end{gathered}  \right) \hfill
\end{gathered} \]
}
When compared, for instance, with the cross-section given in \cite{Parsamyan:2007ju}-\cite{Parsamyan:2010se}, in which the effects due to the $P_T$ to $S_T$ transition have been neglected, in this expression there are new $sin\theta$-scaled terms and $\theta$-depending factors and two extra azimuthal modulations ($\sin \left( {2{\varphi _h} + {\varphi _S}} \right)$ and $\cos \left( {{\varphi _h} + {\varphi _S}} \right)$).

\begin{figure}[h]
\center
\includegraphics[width=0.725\textwidth, angle=-90]{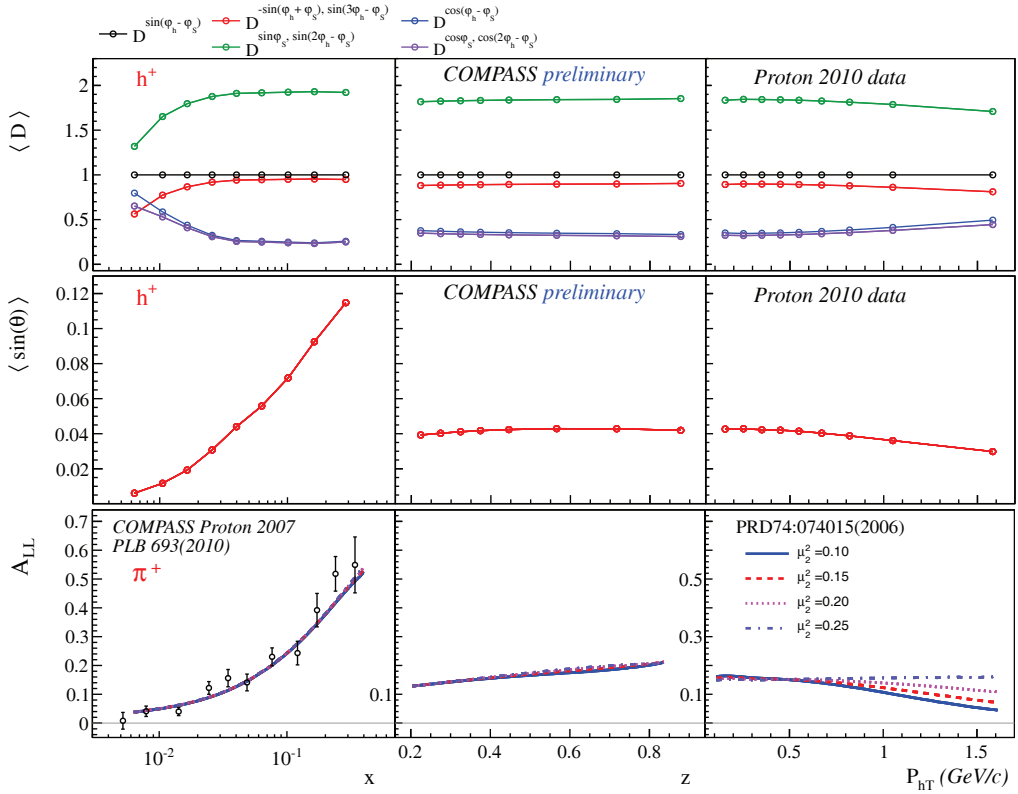}
\caption{Mean depolarization factors $D$ (top), $sin(\theta)$ factors (middle) and $A_{LL}$ asymmetry (bottom).}
\label{fig:DySinTHA_LL}
\end{figure}

The equation (\ref{eq:x_sec_mod}) counts in total eight: five Single-Spin (SSA) and three Double-Spin (DSA) target transverse polarization dependent asymmetries. Since $sin\theta$ is a rather small quantity in COMPASS kinematics (see Fig.\ref{fig:DySinTHA_LL}) the influence of the additional terms, represented by $sin\theta$-scaled longitudinal-spin amplitudes and $\theta$-angle dependent factors, is sizable only in the case of $A_{LT}^{\cos {\varphi _S}}$ DSA, which, even taking into account the smallness of $sin\theta$, is still sizably affected by a large $A_{LL}$ amplitude \cite{Alekseev:2010ub}.
To correct the $A_{LT}^{\cos {\varphi _S}}$ asymmetry we have used the $A_{LL}$ values evaluated based on \cite{Anselmino:2006yc}). Corresponding $A_{LL}$ curves are shown in Fig.\ref{fig:DySinTHA_LL} compared with the COMPASS data points \cite{Alekseev:2010ub}, demonstrating close agreement.

Other transverse amplitudes (including Collins and Sivers) are in much more favorable situation, since they either are not affected by any longitudinal amplitude ($A_{UT}^{\sin (3\phi _h -\phi _s)}$, $A_{UT}^{\sin (\phi _s )}$ and $A_{LT}^{\cos (2\phi _h -\phi _s )}$), or are affected by $A_{UL}^{\sin (\phi _h )}$, $A_{UL}^{\sin (2\phi _h )}$ or $A_{LL}^{\cos (\phi _h )}$ amplitudes which appear to be small at COMPASS with deuteron target \cite{Alekseev:2010dm} and are estimated to be of order of few percent with proton target (see, for instance, predictions for $A_{LL}^{\cos (\phi _h )}$ at COMPASS in \cite{Anselmino:2006yc}).
The effect of possible corrections has been studied on the example of $A_{LT}^{\cos (\phi _h -\phi _s )}$ asymmetry which was corrected for the $A_{LL}^{\cos (\phi _h )}$-contribution from \cite{Anselmino:2006yc}). It was demonstrated that the effect from longitudinal amplitude contribution is negligible.
The eight target transverse spin dependent "raw" asymmetries are extracted simultaneously, using an unbinned maximum likelihood method and then are corrected for the $D$ depolarization factors ($\varepsilon$-depending factors in equation (\ref{eq:x_sec_mod}) standing in front of the amplitudes), dilution factor and target and beam (only DSAs) polarizations \cite{Parsamyan:2007ju},\cite{Parsamyan:2010se}. Obtained from COMPASS data mean values of the $D$ factors corresponding to different asymmetries are presented in Fig.\ref{fig:DySinTHA_LL}.

In the QCD parton model approach four of the eight transverse spin asymmetries ($A_{UT}^{sin(\phi_h+\phi_S)}$, $A_{UT}^{sin(\phi_h+\phi_S)}$, $A_{UT}^{\sin (3\phi _h -\phi _s )}$ SSAs and $A_{LT}^{\cos (\phi _h -\phi _s )}$ DSA) have Leading Order (twist) (LO) interpretation and are described by the convolutions of the twist-two TMD PDFs and FFs, while the other four ($A_{UT}^{\sin (\phi _s )}$ and $A_{UT}^{\sin (2\phi _h -\phi _s )}$ SSAs and $A_{LT}^{\cos (\phi _s)}$ and $A_{LT}^{\cos (2\phi _h -\phi _s )}$ DSAs), despite their higher-twist origin, however, can be represented as "Cahn kinematic corrections" to twist-two effects.
These sub-leading amplitudes are suppressed with respect to the leading twist ones by $\sim M/Q$ (for details see: \cite{Bacchetta:2006tn},\cite{Mulders:1995dh},\cite{Parsamyan:2007ju}-\cite{Parsamyan:2010se}).
It can be shown that LO $A_{UT}^{\sin (3\phi _h -\phi _s )}$ (related to the $h^{q\perp}_{1T}$ "pretzelosity" PDF) is expected to scale according to $\sim|\bf{P_{hT}}^3|$ and thus is suppressed by $\sim|\bf{P_{hT}}|^2$ w.r.t $\sim|\bf{P_{hT}}|$-scaled Collins, Sivers and $A_{LT}^{\cos (\phi _h -\phi _s )}$ LO amplitudes. Similarly, the other four asymmetries are suppressed by $\sim|\bf{P_{hT}}|$.
The $A_{LT}^{\cos (\phi _h -\phi _s )}$ amplitude (related to the $g^{q\perp}_{1T}$ "worm gear" PDF) is of particular interest because it is the only transverse DSA expected to be sizable (LO, no suppression).

For Collins and Sivers effects, COMPASS has recently published results from 2010 proton data \cite{Adolph:2012sn},\cite{Adolph:2012sp}. In the next section we present the preliminary results for the other six asymmetries obtained from the same data sample.
\section{Data analysis and results}
\begin{figure}[H]
\center
\includegraphics[width=0.95\textwidth]{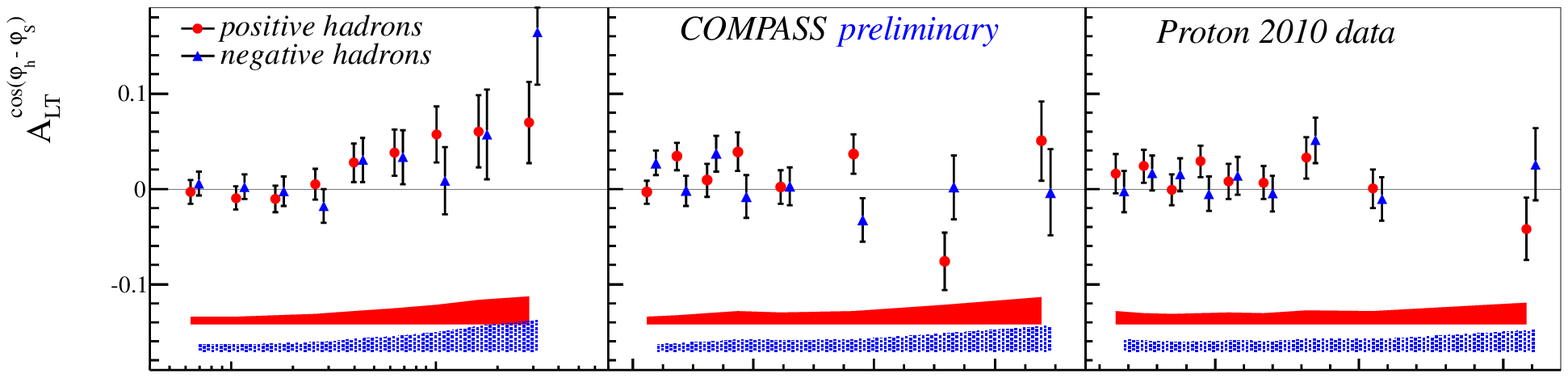}
\includegraphics[width=0.95\textwidth]{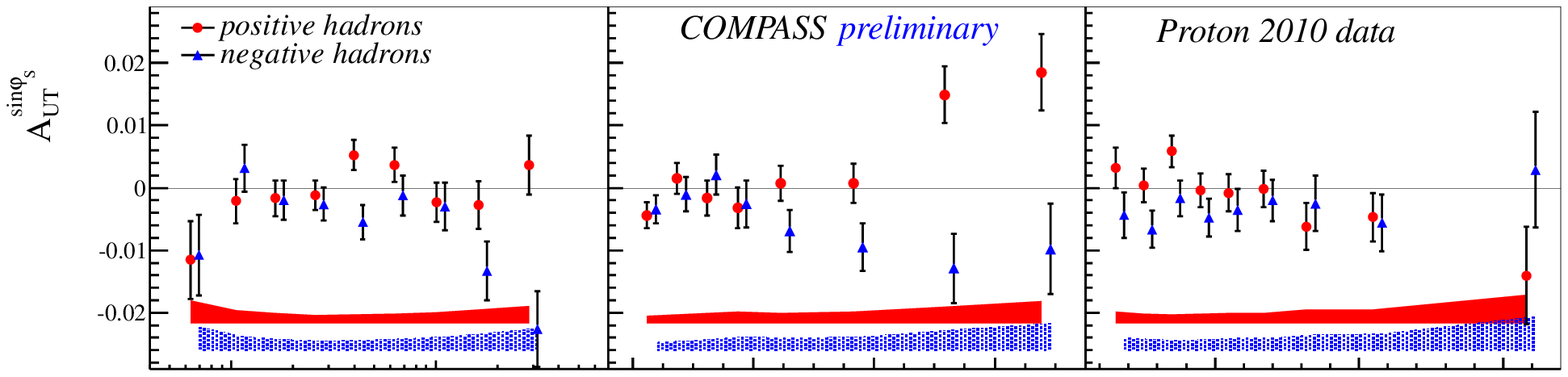}
\includegraphics[width=0.95\textwidth]{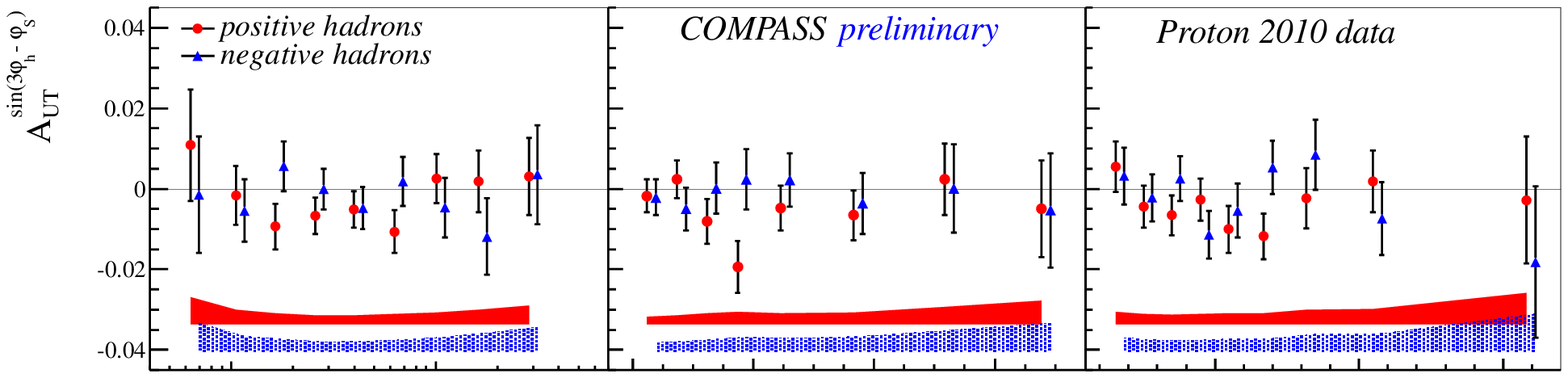}
\includegraphics[width=0.95\textwidth]{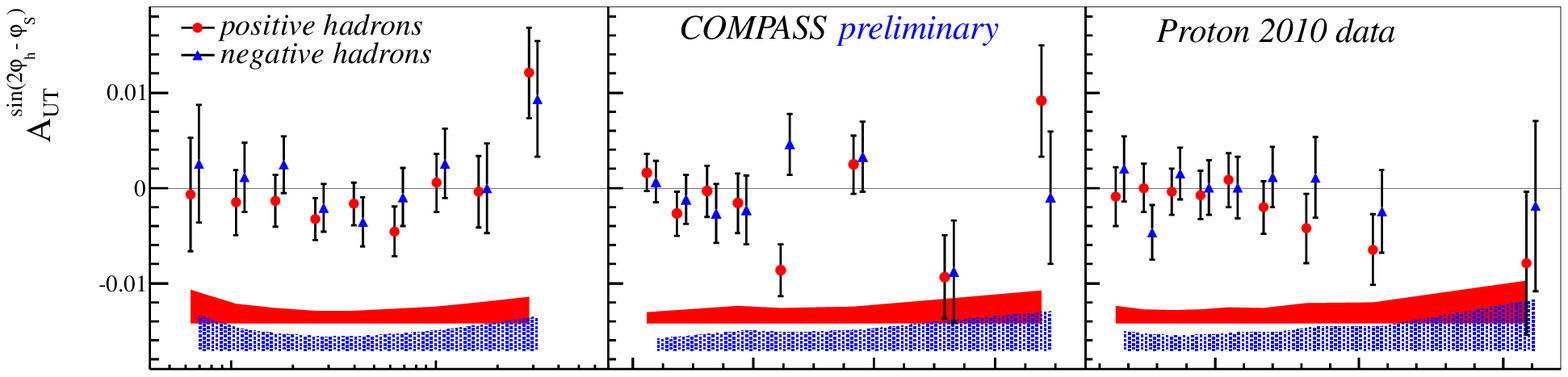}
\includegraphics[width=0.95\textwidth]{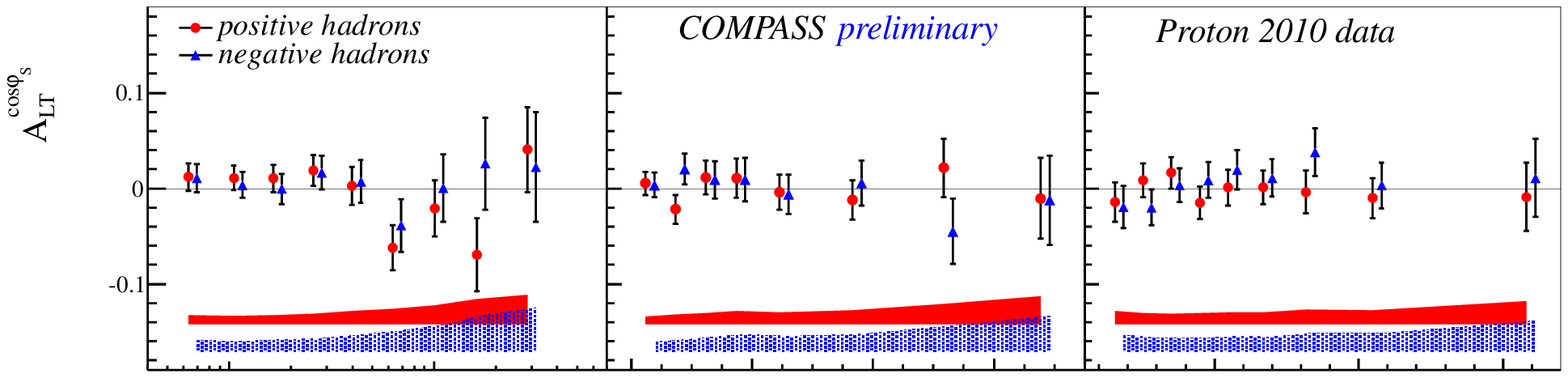}
\includegraphics[width=0.95\textwidth]{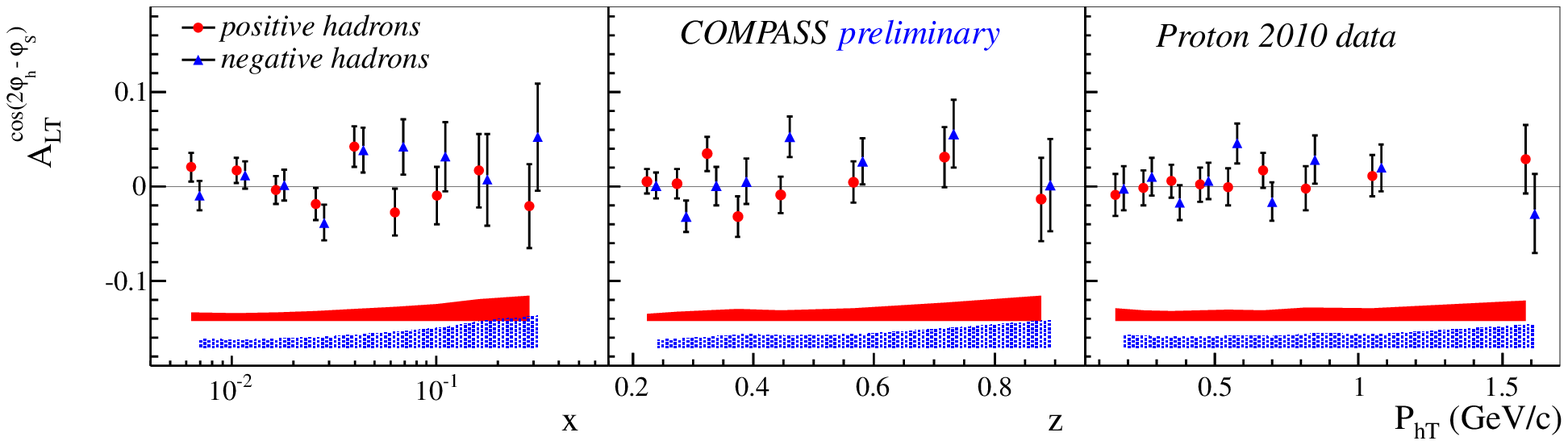}
\caption{Six "Beyond Collins and Sivers" asymmetries at COMPASS (Proton 2010 data).}
\label{fig:A8}
\end{figure}

\begin{figure}[h]
\center
\includegraphics[width=0.95\textwidth]{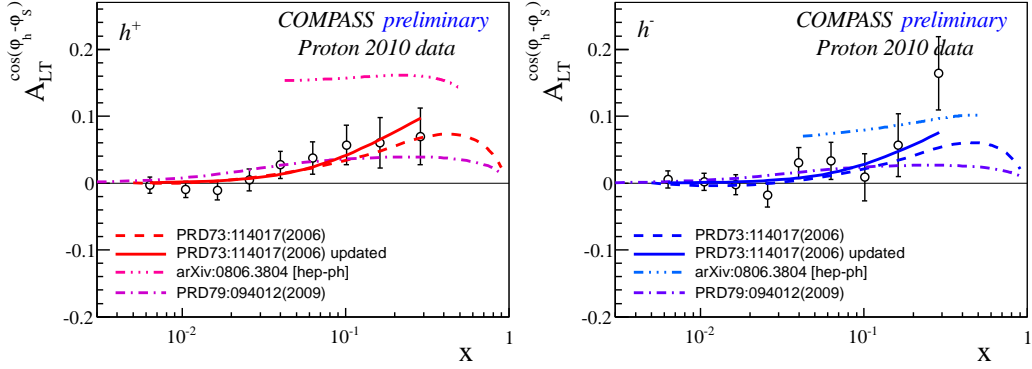}
\caption{$A_{LT}^{cos(\phi_h-\phi_S)}$ asymmetry: comparison with the theoretical predictions.}
\label{fig:A_LT}
\end{figure}

The whole data selection and analysis procedure applied for the extraction of the six mentioned asymmetries from COMPASS 2010 proton data
is identical to those applied for the already published Collins and Sivers asymmetries.
The detailed description of COMPASS spectrometer and details on analysis can be found in: \cite{Abbon:2007pq},\cite{Parsamyan:2007ju},\cite{Parsamyan:2010se},\cite{Adolph:2012sn},\cite{Adolph:2012sp} (and references therein).
%
%The kinematic cuts applied in the analysis are the following ones:
%$Q^2>1$ (GeV/c)$^2$, $W>5$ GeV, $0.1<y<0.9$, $P_{hT}>0.1$ GeV/c and
%$z>0.2$.
%
The asymmetries extracted as functions of $x$, $z$ and $P_{hT}$ for positive and negative hadrons are shown in Fig.\ref{fig:A8}. The systematic uncertainties for each asymmetry have been estimated separately for positive and negative hadrons and are given by the bands.
According to the preliminary observations, there is an evidence of non-zero LO $A_{LT}^{\cos (\phi _h -\phi _s )}$ DSA and sub-leading $A_{UT}^{\sin (\phi _s )}$ SSA , while the other four "beyond Collins and Sivers" amplitudes are found to be compatible with zero within the statistical accuracy. It has to be mentioned that similar behavior for both non-zero amplitudes and for zero-compatible amplitudes was already preliminary reported also by the HERMES collaboration.
In Fig.\ref{fig:A_LT} $A_{LT}^{\cos (\phi _h -\phi _s )}$ asymmetry, extracted from COMPASS 2010 proton data, is compared with the theoretical predictions from \cite{Kotzinian:2006dw}, \cite{Boffi:2009sh} and \cite{Kotzinian:2008fe}, demonstrating a good level of agreement between theory and measurement within the given statistical accuracy.
%and provide an evidence for the worm-gear distribution.
%All the obtained results will be the subject of a future publication.
%
%
\section{Conclusions}\label{concl}
The preliminary results on six, additional to Collins and Sivers amplitudes, asymmetries from COMPASS proton 2010 data, have been presented. A non-zero trend has been observed for the $A_{LT}^{\cos (\phi _h -\phi _s )}$ and $A_{UT}^{\sin (\phi _s )}$ amplitudes, while the other four are found to be consistent with zero within the statistical accuracy. The measured kinematical dependencies of $A_{LT}^{\cos (\phi _h -\phi _s )}$ asymmetry are inline with the predictions given by several theoretical models.
Combined with the previous COMPASS measurements and data from other experiments, these results give another possibility to access TMD PDFs and FFs, and to study the spin-structure of the nucleon.

\end{document}